\documentclass{article}
\usepackage{spconf,amsmath,epsfig,amssymb}

\usepackage{graphicx}

\usepackage{xcolor}

\usepackage{tabularx}
\usepackage{arydshln}
\usepackage{authblk}

\title{A cycle GAN approach for Heterogeneous Domain Adaptation in Land Use classification}

\threeauthors{Claire Voreiter$^{1}$, Jean-Christophe Burnel$^{1}$, Pierre Lassalle$^{2}$,}{Marc Spigai$^{3}$,}{Romain Hugues$^{3}$, Nicolas Courty$^{1}$}{$^{1}$Universit\'{e} de Bretagne Sud - IRISA, Rue Yves Mainguy, 56000 Vannes - France}{$^{2}$CNES, 18 avenue Edouard Belin, 31400 Toulouse Cedex 9 - France}{$^{3}$Thales Alenia Space, 26 Avenue Jean Fran{\c c}ois Champollion, 31100 Toulouse - France}

\begin{document}

\maketitle

\begin{abstract}
In the field of remote sensing and more specifically in Earth Observation, new data are available every day, coming from different sensors. Leveraging on those data in classification tasks comes at the price of intense labelling tasks that are not realistic in operational settings. While domain adaptation could be useful to counterbalance this problem, most of the usual methods assume that the data to adapt are comparable (they belong to the same metric space), which is not the case when multiple sensors are at stake. Heterogeneous domain adaptation methods are a particular solution to this problem. We present a novel method to deal with such cases, based on a modified cycleGAN version that incorporates classification losses and a metric space alignment term. We demonstrate its power on a land use classification tasks, with images from both Google Earth and Sentinel-2. 
%In the field of remote sensing and more specifically in Earth Observation, new data are available every day, coming from different sensors. It is expensive to labeled them and take advantage of this data through domain adaptation allow to reduce this cost. Each sensor has is own characteristics and images do not evolve in the same domain according to their source. We are then in an heterogeneous domain adaptation problem. A method to deal with and to generate different images is Cycle GAN. In our case, we have source images coming from Google Earth and target images coming from Sentinel-2. Our work takes advantage from Cycle GAN to generate images and target labels from the source domain in which we have all the labels. Then, we do a Land Use classification task on the target data thanks to the generated data. \\
\end{abstract}


\section{Introduction}
In remote sensing, a lot of data are available but disposing of associated labels is usually hard and time-consuming. Furthermore, related to the sensor from which the data come from, they do not share the same characteristics. While leveraging on one labelled dataset (the source domain) to perform classification on a new dataset (target) with similar characteristics can be cast as a domain adaptation problem, handling data sources from different sensors is more tricky. This problem is known as an heterogeneous domain adaptation task. In this case, the data have different characteristics (that can include spatial or spectral resolutions) and evolve in different domains. Heterogeneous domain adaptation (HDA) is usually divided in two categories : semi-supervised HDA, when a (small) quantity of target labels are available, and unsupervised HDA, when no target labels at all is available.
The difficulty of heterogeneous domain adaptation is to find a sensitive way of linking together the two different domains. 

In the state of the art, methods dealing with HDA are mostly shallow methods. EGW and SGW~\cite{yan2018semi} aim to learn an optimal coupling between the two domains through a Gromov-Wasserstein distance. SHFA~\cite{li2013learning} aims at augmenting source and target samples based on two projection matrices and train simultaneously an SVM classifier on the augmented data. CDLS~\cite{hubert2016learning} aims to find representative landmarks to learn a domain-invariant feature subspace and then train a classifier in the learned subspace for target data. DCA~\cite{yan2017learning} aims to jointly obtain a discriminative correlation subspace defined by CCA and then learn a classifier in this subspace. Those methods are not directly amenable to deep learning and the related import quantity of data.

Our new approach for HDA, that is well suited in a deep learning context, is based on the concept of CycleGAN~\cite{cherian2019sem,zhu2017unpaired,russo2018source}. This method is based on the notion of Generative Advsersarial Networks (GAN)~\cite{goodfellow2014generative} which consists in putting two networks in competition: a generator network that produces data and a discriminator network which tries to distinguish the generated sample from a true data.
The core idea of CycleGAN is to find an invertible mapping between two domains where examples are provided. Contrary to GANs, no source of noise is considered, two generators (functions that implement mappings to and from the domains are considered, and two discriminators for both domains. We cast the HDA in this framework, but since we want to perform a classification task we need to preserve labels during the image generation. Our contribution focuses on this issue. With the help of this new labeled images, we achieve a classification task on the target domain. A few set of labeled data are used from the target domain and we aim to reduce as much as possible the number of available target labels. 

\section{Proposed method}

Two sets of data are used, source data $X_s = \{x_i^s\}_{i=1}^{N_s}$ with $\mathbb{P}_s$ the data distribution, in the domain $S \in \mathbb{R}^{m \times n \times d}$, with $y_s = \{y_i^s\}_{i=1}^{N_s}$ the class labels associated, and, target data $X_t = \{x_i^t\}_{i=1}^{N_t}$ with $\mathbb{P}_t$ the data distribution, in the domain $T \in \mathbb{R}^{k \times l \times e}$, with $n_{y_t}$ labels $y_t = \{y_i^t\}_{i=1}^{N_t}$. Our method is based on the CycleGAN idea and the notion of GAN. We are going to present this two methods before introducing our improvements.

\subsection{GAN}

The idea of GANs \cite{goodfellow2014generative} consist in putting in competition two networks. A generator network $G$ maps a noise and generate samples, and, a discriminator $D$ network tries to distinguish the generated sample from true data samples. The loss function is a minimax game between the generator and the discriminator.
We have $x$ the real data, $\mathbb{P}_r$ the data distribution, $\tilde{x} = G(z)$ with $z$ the input of the generator, sampled from some a noise distribution and $\mathbb{P}_G$ the the model distribution.
The loss function $\mathcal{L}_{GAN}$ of a GAN, which is minimized over $G$ and maximized over $D$, reads:
$$\displaystyle \mathcal{L}_{GAN}(G,D)=\underset{x \sim \mathbb{P}_r}{\mathbb{E}} [\log(D(x))] + \underset{\tilde{x} \sim \mathbb{P}_G}{\mathbb{E}} [\log(1-D(\tilde{x}))] .$$

The CycleGAN method lies in this idea with the addition of a cycle consistency term during the generation.

\subsection{Cycle GAN}

The Cycle GAN \cite{zhu2017unpaired} is a method based on GAN used to perform style transfer. The major advantages of this methods are the weakly supervision and the fact that it does not need paired images. The goal of Cycle GAN is to learn mapping function between the two domains $S$ and $T$. To achieve this, we have two generators, $G_{s2t} : S \rightarrow T$ and $G_{t2s} : T \rightarrow S$, and two discriminators, $D_s$ which aims to distinguish source images from images generated by $G_{t2s}$ and $D_t$ which aims to distinguish source images from images generated by $G_{s2t}$. The Cycle GAN statement lies in the cycle consistency idea. It is based on the fact that the generated data should be re-transform to their original form: $G_{t2s}(G_{s2t}(x^s)) \approx x^s$ and $G_{s2t}(G_{t2s}(x^t)) \approx x^t$
The cycle consistency loss measures the difference between the original
images and the two times generated images. This loss can be written as : 
\begin{eqnarray*}
\displaystyle \mathcal{L}_{cycle} (G_{s2t}, G_{t2s}) =  \mathbb{E}_{x^s \sim \mathbb{P}_s} ||x^s - G_{t2s}(G_{s2t}(x^s)|| +  \\
\displaystyle \mathbb{E}_{x^t \sim \mathbb{P}_t} ||x^t - G_{s2t}(G_{t2s}(x^t)|| ,
\end{eqnarray*}
where $||$ . $||$ is a suitable norm. With the cycle consistency loss, an adversarial loss is applied for both discriminators. The global loss is then :
\begin{eqnarray*}
\displaystyle
\mathcal{L}_{cycleGAN}(G_{s2t}, G_{t2s}, D_s, D_t) = \mathcal{L}_{GAN}(G_{s2t}, D_t) \\
\displaystyle + \mathcal{L}_{GAN}(G_{t2s}, D_s) + \lambda \mathcal{L}_{cycle}(G_{s2t}, G_{t2s}) .
\end{eqnarray*}

\subsection{Proposed architecture}
The Cycle GAN method can be apply for our case to generate source images in the target domain or the other way around. However, because we are handling a classification problem, specific cares have to be taken to ensure a correct transfer of the labels.
%\nc{However, being able to map correctly two samples in the two domains does not guarantee that nearby samples in one domain will be mapped close to each other in the other domain. This issue is linked to the notion of regularity of the mapping $G$, which is for example described in~\cite{}. Our method uses two extra classifiers and a metric space consistency term to handle this constraint, and help in transferring labels.}

\subsubsection{Metric alignment, classification and total losses}
With a Cycle GAN architecture, two samples can be arbitrarily mapped in the target space and it is difficult to ensure that two nearby samples in the target domain will share a correct label. Yet, most classifiers are using this regularity to enforce generalization. This issue is linked to the notion of regularity of the mapping $G$, which is for example described in~\cite{UDT2019}.
We add a metric alignment loss that will help us to assure that two close samples from the source domain must be close in the target domain after the generation, and the other way around. The corresponding loss simply reads:
$$ \displaystyle \mathcal{L}_{metric}(G) = \mathbb{E}_{(x_i, x_j) \sim \mathbb{P}_s, } (d(x_i,x_j)-d(G(x_i),G(x_j)))^2 ,$$
where $d$ is a distance between the samples in the corresponding domains. We notably simply choose a simple Euclidean distance, but this choice could be improved in the future.

This classification consistency is also ensured by a specific  loss based on the output of the classifiers $C$:
$$ \displaystyle \mathcal{L}_{classif}(C,G) = \mathbb{E}_{(x,y)} l(C(G(x)),y),$$
evaluated over labelled samples from both source and target domains (if any), and where $l(.,.)$ is a classification loss (typically cross-entropy in a deep learning setting) that forces the sample coming from the generators to be classified according the true associated labels.

Finally, the global loss of our method is: 
\begin{align*} &\displaystyle \mathcal{L}(G_{s2t}, G_{t2s}, D_s, D_t) =\\
&\mathcal{L}_{cycleGAN}(G_{s2t}, G_{t2s}, D_s, D_t) + \mathcal{L}_{classif}(C_s) \\ 
\displaystyle &+ \mathcal{L}_{classif}(C_t) + \mathcal{L}_{metric}(G_{s2t}) + \mathcal{L}_{metric}(G_{t2s}).
\end{align*}

\begin{figure}
\centering
\includegraphics[width=\columnwidth]{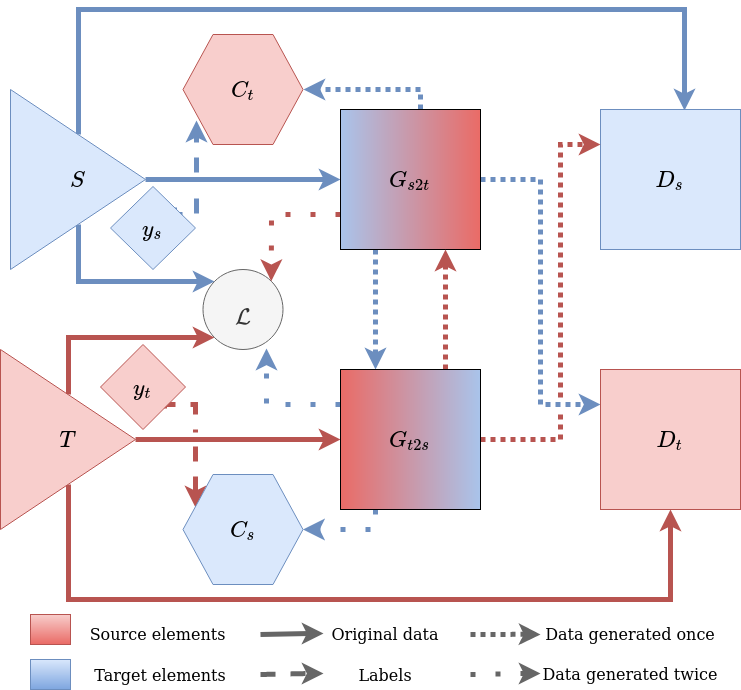}
\caption{Diagram of our architecture showing the different elements of its composition.}\label{architecture}
\end{figure}

\subsubsection{Overall architecture}

In our architecture (see Fig.~\ref{architecture}), we recover the elements of the Cycle GAN plus two classifiers $C_s$ and $C_t$. 
We pretrain $C_s$ with the source data and their labels. Then, we use it to test the relevance of the generated images coming from $G_{t2s}$ with the help of the available target labels.
In the same way, we pretrain $C_t$ with the target images and the labels available (when any), and we use it to test the relevance of the generated images coming from $G_{s2t}$ with the help of the source labels. 
Final classification on the target data is achieved with a third classifier which is trained independently from the architecture of Fig.~\ref{architecture}, and discussed subsequently.

%\begin{algorithm}
%\caption{Algorithm}
%\hspace*{\algorithmicindent} \textbf{Input} source data $X_s = \{x_i^s\}_{i=1}^{N_s}$ with associated labels $y_s = \{y_i^s\}_{i=1}^{N_s}$; target data $X_t = \{x_i^t\}_{i=1}^{N_t}$ with $n_{y_t}$ associated available labels $y_t = \{y_i^t\}_{i=1}^{N_t}$
%
%\begin{algorithmic}
%\STATE Pretrain $C_s$ with $x_s$ and $y_s$
%\STATE Pretrain $C_t$ with $x_t$ and $y_t$
%\WHILE {$i <$ iteration number }
%\STATE $G_{s2t}(x^s) \leftarrow x^s$
%\STATE $D_s(x^s) \leftarrow x^s$
%\STATE $G_{t2s}(G_{s2t}(x^s)) \leftarrow G_{s2t}(x^s)$
%\STATE $D_t(G_{s2t}(x^s)) \leftarrow G_{s2t}(x^s)$
%\STATE Test $G_{s2t}(x^s)$ with $C_t$ and associated $y_s$
%\STATE $G_{t2s}(x^t) \leftarrow x^t$
%\STATE $D_t(x^t) \leftarrow x^t$
%\STATE $G_{s2t}(G_{t2s}(x^t)) \leftarrow G_{t2s}(x^t)$
%\STATE $D_s(G_{t2s}(x^t)) \leftarrow G_{t2s}(x^t)$
%\STATE Test $G_{t2s}(x^t)$ with $C_s$ and associated $y_t$
%\STATE Compute loss function
%\STATE $i \leftarrow i+1$
%\ENDWHILE
%
%\end{algorithmic}
%\end{algorithm}

\subsubsection{Final classification}

\begin{figure*} 
 \begin{minipage}[b]{.22\linewidth} 
  \centering\epsfig{figure=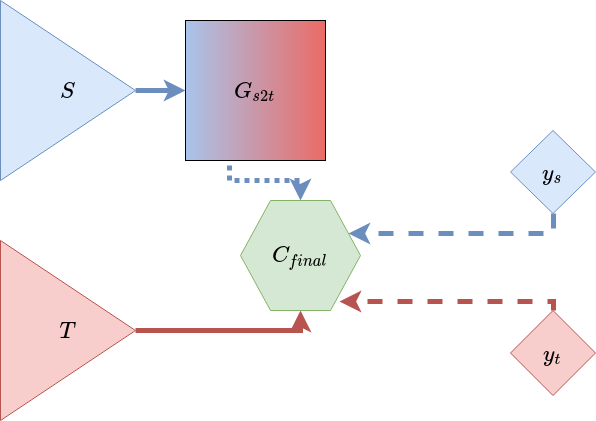,width=\linewidth} 
 \end{minipage} \hfill 
 \begin{minipage}[b]{.32\linewidth} 
  \centering\epsfig{figure=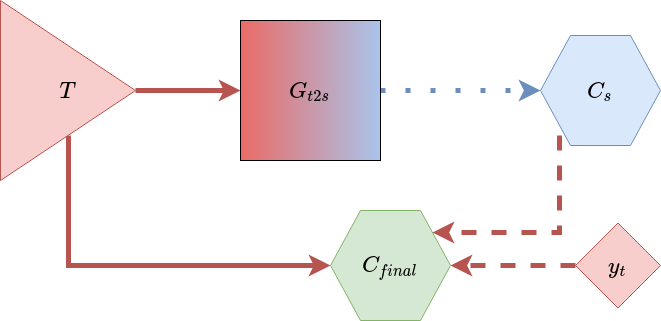,width=\linewidth} 
 \end{minipage} \hfill
 \begin{minipage}[b]{.27\linewidth} 
  \centering\epsfig{figure=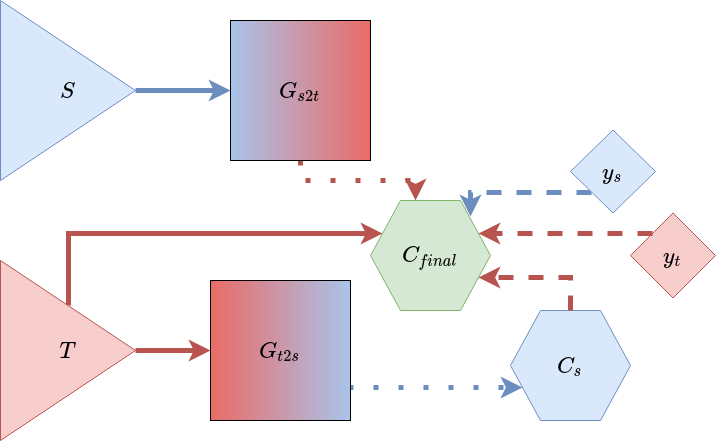,width=\linewidth} 
 \end{minipage} 
 \caption{On the left, diagram of the {\tt HDAsource} method. In the middle, diagram of the {\tt HDAtarget} method. On the right, diagram of the {\tt HDAfull} method. This diagrams are showing the origin of the data use for the final classifier. }
 \label{schema}
\end{figure*}

Once the transformation architecture has been trained, there are three ways of performing classification of the unlabelled target samples. In all cases, we use the labelled target data (if any), and we add: {\em i)} source images transferred to the target domain by $G_{s2t}$ and their corresponding labels. 
In the remainder, we refer to this method as {\tt HDAsource}; {\em ii)} unlabelled target images, associated to the label obtained by transforming them into their source equivalent (via $G_{t2s}$) and associated to a label obtained by $C_s$ ({\tt HDAtarget});
{\em iii)} a combination of both approaches to generate learning data ({\tt HDAfull}). Fig.~\ref{schema} depicts the three approaches for training the final classifier.

\section{Experiments and Results}

To assess the qualities of the proposed method, we chose two datasets to make Land Use classification. We selected NWPU-RESISC45 and EuroSAT datasets.

\subsection{Datasets}

NWPU-RESISC45 \cite{cheng2017remote} contains 31,500 images and cover 45 scene classes. The images come from Google Earth, have a size of 256*256 and are in RGB. The spatial resolution of this dataset varies from about 30 m to 0.2 m per pixel. 
EuroSAT \cite{helber2017eurosat}, \cite{helber2018introducing} contains 27,000 images and cover 10 scene classes. The images come from Sentinel-2 satellite, are composed by 13 spectral bands and have a size of 64*64. 
Since EuroSAT dataset contains only 10 classes, we choose to start from this dataset, we look for common classes between the two datasets. We have 9 classes (annual crop, permanent crop, forest, highway, industrial, pasture, residential, river and sea lake) and we choose to merge annual crop and permanent crop into crop because the images were very similar. Then we selected the corresponding one from NWPU-RESISC45 dataset. For each classes, 700 labelled samples are available. We take 650 samples at best to test our architecture and we keep 50 samples to do the validation.

\subsection{Results}

\begin{figure}
\centering
\includegraphics[width=1\columnwidth]{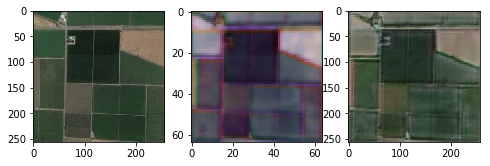}
\includegraphics[width=1\columnwidth]{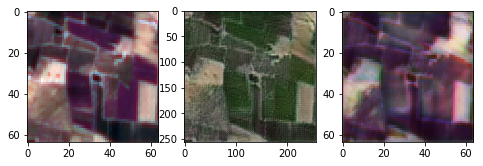}
\caption{On the left, we have the original images, in the center, we have the first generated images, and on the right, we have the second generated images which should look like the original images.}
\label{crop_ts}
\end{figure}

We show in Fig.~\ref{crop_ts} images generated by our method.
We want to analyse the influence of the number $n_{y_t}$ of target labels available, so we run our tests with different amount of target labels. 

For the model we use a similar architecture as SRGAN \cite{ledig2017photo} for the generators, a convNet for the discriminators made up of four convolutional layers, with batch normalisation and ReLU activation function and also a convNet for the two classifiers made up of four convolutional layers with dropout, maxpooling and ReLU activation function. The final classifier trained is also a convNet made up of three convolutional layers with ReLU activation function. We train the model for 300.000 iterations and the final classifier on 30 epochs.

\begin{table}
    \begin{center}
     \begin{tabularx}{8.6cm}{ X | X X X X }
     
       \centering{\# $n_{y_t}$} & \centering{{\footnotesize\tt baseline}} & \centering{{\footnotesize\tt HDAsource}} & \centering{{\footnotesize\tt HDAtarget}} & \centering{{\footnotesize\tt HDAfull}} \tabularnewline
       \hline
       \centering 650  & \centering 77.00 & \centering 71.25 & \centering \textbf{80.00} & \centering 75.00 \tabularnewline
       \hdashline
       \centering 325 & \centering 70.75 & \centering 70.00 & \centering \textbf{77.00} & \centering 72.00 \tabularnewline
       
       \centering 130 & \centering 64.00 & \centering 50.75 & \centering \textbf{70.00} & \centering 66.50 \tabularnewline
       
       \centering 65 & \centering 62.00 & \centering 43.00 & \centering \textbf{68.50} & \centering 64.25 \tabularnewline
       
       \centering 1 & \centering 19.00 & \centering 32.25 & \centering \textbf{66.75} & \centering 62.00 \tabularnewline
       \hdashline
       \centering 0 & \centering  - & \centering 30.75 & \centering \textbf{65.75} & 57.50\centering \tabularnewline
       
     \end{tabularx}
    \end{center}
   \caption{Rate of accuracy scores for classification on EuroSAT dataset}
   \label{tab1}
\end{table}

We compare the accuracy scores of 4 methods: {\footnotesize\tt baseline}, that is the score when only labelled samples in the target are considered, {\footnotesize\tt HDAsource}, {\footnotesize\tt HDAtarget} and {\footnotesize\tt HDAfull}. As one can observe from the results of table \ref{tab1}, our method gives better classification results than with target labeled data only in all cases. Less we have available target labels, more we improve the classification results. Interestingly enough, {\footnotesize\tt HDAtarget} gives the best score consistently. This can be explained by the following fact: since labels are available in the source domain, the $C_s$ classifier is very good, and helps in training a performant $G_{t2s}$ transformation. A contrario, the transformation from source to target is less performant, and when combined in the {\footnotesize\tt HDAfull} setting, it only worsen the classification accuracy. Also, the best performance 66\% in the fully unsupervised setting  is very good (only 11 points behind the golden score (77\%), which indicates that our method is capable of transferring knowledge even in the absence of labelled samples in the target domain.

\section{Conclusions}

In this paper, we presented a novel approach for classification with heterogeneous domain adaptation using a cycle GAN based approach. Our contribution lies in the addition of specific classification and metric alignment losses, that also helps in the generation process. Experimental results on a Land Use classification problem involving very different remote sensing images indicate the power of our method, in both semi-supervised and unsupervised settings. Future works will consider ablation studies to fully undertstand the role of each loss independently, and compare with existing unsupervised or semi-supervised deep HDA methods.

\section{Acknowledgement}

This work was supported by Centre National d'\'{E}tudes Spatiales (CNES) and Thales Alenia Space.

\footnotesize
\bibliographystyle{IEEEbib}
\bibliography{source}

\end{document}